\documentclass{article}
\newcommand{\vek}[1]{ {\bf #1}}
\usepackage{epsfig}

\begin{document}

\begin{center}
\textbf{THE 2005 - 2010 MULTIWAVELENGTH CAMPAIGN OF OJ287}
\end{center}

\begin{center}
\textbf{Mauri Valtonen$^{1,2,3}$, Aimo Sillanp\"a\"a$^2$}
\end{center}

\begin{center}
{\it
\noindent $^1$Helsinki Institute of Physics, University of Helsinki\\
$^2$Department of Physics and Astronomy, University of Turku\\
$^3$Finnish Centre for Astronomy with ESO, University of Turku\\}
\end{center}

\begin{abstract}
The light curve of quasar OJ287 extends from 1891 up today without major gaps. This is partly due to extensive studies of historical plate archives by Rene Hudec and associates, and partly due to several observing campaigns in recent times. Here we summarize the results of the 2005 - 2010 observing campaign, in which several hundred scientists and amateur astronomers took part. The main results are the following: (1) The 2005 October optical outburst came at the expected time, thus confirming the General Relativistic precession in the binary black hole system, as was originally proposed by Sillanp\"a\"a et al. (1988). At the same time, this result disproved the model of a single black hole system with accretion disk oscillations, as well as several toy models of binaries without relativistic precession. In the latter models the main outburst would have been a year later. No particular activity was seen in OJ287 in 2006 October. (2) The nature of the radiation of the 2005 October outburst was expected to be bremsstrahlung from hot gas at a temperature of $3\times 10^{5~\circ}$K. The reason for the outburst is a collision of the secondary on the accretion disk of the primary, which heats the gas to this temperature. This was confirmed by combined ground based and ultraviolet observations using the XMM-Newton X-ray telescope. (3) A secondary outburst of the same nature was expected at 2007 September 13. Within the accuracy of the observations (about 6 hours), it started at the correct time. Thus the prediction was accurate at the same level as the prediction of the return of Halley's comet in 1986. Due to the bremsstrahlung nature of the outburst, the radiation was unpolarised, as expected. (4) Further synchrotron outbursts were expected following the two bremsstrahlung outbursts. They came as scheduled between 2007 October and 2009 December. (5) Due to the effect of the secondary on the overall direction of the jet, the parsec scale jet was expected to rotate in the sky by a large angle around 2009. This rotation has been seen in high frequency radio observations. The OJ287 binary black hole system is currently our best laboratory for testing theories of gravitation. Using OJ287, the correctness of General Relativity has now been demonstrated up to second Post-Newtonian order, higher than has been possible using binary pulsars.\\
\\
\noindent \textbf{Keywords}: quasars: general - quasars: individual (OJ287) - BL Lacertae objects: individual (OJ287)
\end{abstract}

\section{Introduction}

OJ287 is one of the brightest AGN in the sky. Since it is also highly variable, it has become one of the favorite objects for both professional and amateur astronomers to follow. In addition, it lies close to the ecliptic, which means that its image has been recorded by chance in hundreds of photographic plates since 1891.
 
In 1982 one of the authors (A.S.) put together the historical light curve of OJ287 based on published measurements. These were partly photometric measurements since the discovery of OJ287 as an extragalactic object in 1968, partly studies of photographic plate archives from years prior to 1968 that are kept in various observatories, in particular at Harvard and at Sonneberg. There appeared to be a 12 year outburst cycle (see Figure 1), and moreover, it was obvious that the next cyclic outburst was due very shortly. This prediction was distributed to colleages world wide, and indeed, OJ287 did not disappoint us but produced the expected event in the following January [1,2]. Observations showed a sharp decline in the percentage polarization during the outburst maximum, indicating that the outburst was produced essentially by unpolarized light [3]. This is different from ordinary outbursts in OJ287 which are characterized by an increase in the percentage polarization at the maximum light. In radio wavelengths the outbursts were found to follow the optical outbursts with a time delay of between 2 months and a year, depending on the observing frequency [Ref. 4].

\begin{figure}
\includegraphics[width=5cm,angle=270]{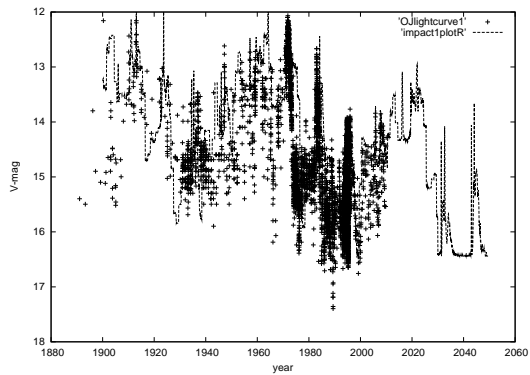} 
 \caption{The optical light curve of OJ287 from 1891 to 2010. The observations are taken from Ref. 22, complemented by unpublished data from R.Hudec and M.Basta}
\end{figure} 

Sillanp\"a\"a et al. [Ref. 5] suggested that OJ287 is a binary black hole system where a smaller companion periodically perturbs the accretion disk of a massive primary black hole. They stated that the best way to verify this hypothesis was to study future outbursts and to show that the major axis of the binary system precesses as expected in General Relativity. The next expected outburst was in 1994; it came as scheduled [Ref. 6, 7]. At this point it became obvious to us that we are indeed dealing with a relativistic binary system, and it became necessary to develop the model in greater detail. In the binary model there should be two disk crossings per 12 yr orbital period. Thus the 1994 outburst should have an equal pair whose timing was calculated to be at the beginning of 1995 October [Ref. 8  - 10]. This prediction was also verified [Ref. 11].

Figure 1 shows the optical light curve from 1891 to 2010. The main gap in the light curve is between years 1893 and 1897; observations from three consecutive years are missing there. There are no major gaps since 1897. (Ref.12 makes a peculiar statement that ``there are about 10 year long gaps in the data''; no such gaps are seen even in their own historical light curve.)

Alternative explanations have also been put forward. Quasiperiodic oscillations in an accretion disk were suggested [Ref. 13], and several binary toy models without relativistic precession have also been proposed [Ref. 14 - 16]. The latter models all predicted the next main outburst of OJ287 in the autumn of 2006, while the precessing binary model gave a prediction one year earlier, at the beginning of 2005 October [Ref. 17, 18]. The second major outburst was expected in late 2007 in all binary models, while in the single black hole model there was no reason to expect a second major outburst. In the precessing binary model the date was given with high accuracy, with the last prediction prior to the actual event being 2007 September 13 [Ref. 19, 20]. In the single accretion disk model and in the non-precessing binary models the nature of the radiation at these outbursts should have been polarized synchrotron radiation, while the precessing binary model predicted unpolarized bremsstrahlung radiation [Ref. 9]. In addition, the precessing binary model predicted a series of further outbursts for the interval 2007 - 2010, but they were expected to show up as an increased level of synchrotron radiation [Ref. 17]. Also, the companion black hole should affect the disk of the primary in a predictable way, leading to the wobble of the jet [Ref. 21]. In contrast, the non-precessing models predicted simultaneous brightening both in radio and in optical, at least for the second outburst [Ref. 16] since in these models disk impacts play no role or a minor role, and flux enhancements are purely jet phenomena.

With these predictions in mind, a multiwavelength campaign of observing OJ287 during 2005 - 2010 was set up, with one of the authors (A.S.) among the leaders.

\section{Five  ``smoking gun'' results}

In the following, we describe five  ``smoking gun'' observations which produced expected results from the point of view of the precessing binary black hole model, but which were surprising and unexpected in other theories.

\subsection{Timing the 2005 outburst}

The 2005 outburst was well covered by observations. The points in Figure 2 are daily averages, 92 in all, formed from altogether 2329 observations in V-band and in R-band. The latter are transformed to V-band by adding 0.4 magnitudes to the R-band value. Finally the flux values are calculated in a standard way (see e.g. Ref. 22).
 
\begin{figure}
\includegraphics[width=5cm,angle=270]{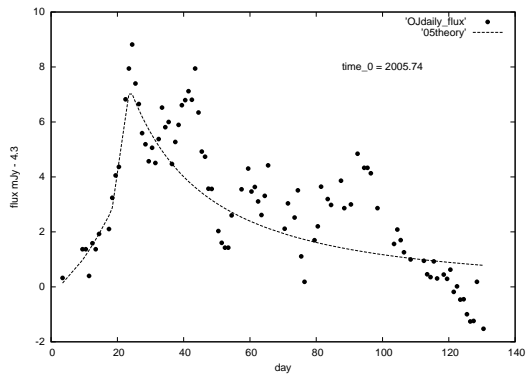} 
 \caption{The optical light curve of OJ287 during the 2005 outburst. The data points are based on Refs. 23, 24 and 25. The dashed line is the theoretical fit based on Ref. 9.}
\end{figure} 

According to Ref. 10, the impact causing the 2005 outburst was expected 22.3 years after the impact of the 1983 outburst. In addition, in Ref. 9 it is estimated that the 2005 outburst should be delayed after the impact. The 1983 outburst is also delayed but not as much, the difference being 0.44 yr. The timing uncertainty was estimated to be $\pm 0.1$ yr. The rapid flux rise started in the latter outburst at 1983.00; thus the corresponding rapid flux rise of the 2005 outburst was expected at 1983.00 + 22.30 + 0.44 = 2005.74. Actually, the outburst was one week late and did not begin until 2005.76 (Ref. 23), but anyway the timing was well within the stated error limits. The comment in Ref. 12 that ``No one expected a major burst at this point'' is rather strange, and it fails to understand that any prediction has its associated error limits. Only a few polarization measurements were carried out at that time, and unfortunately, even those happened during secondary flares. Thus the polarization state of the primary outburst remains unknown from observations. (In contrast, Ref. 12 states that ``the whole burst was rather strongly polarized'', based on two measurements among more than 2000 light curve points, an extraordinary extrapolation!)

Sillanp\"a\"a et al. [Ref. 5] stated that the binary system should show forward precession and thus the disk crossings should follow each other at shorter intervals than the orbital period. The required amount of precession is easily calculated, and it turns out to be $39.1^{\circ}$ per period. It is so much higher than e.g. in  binary pulsars (by a factor of $10^4$) that we immediately realise the importance of OJ287 in testing General Relativity.

\begin{figure}
\includegraphics[width=5cm,angle=270]{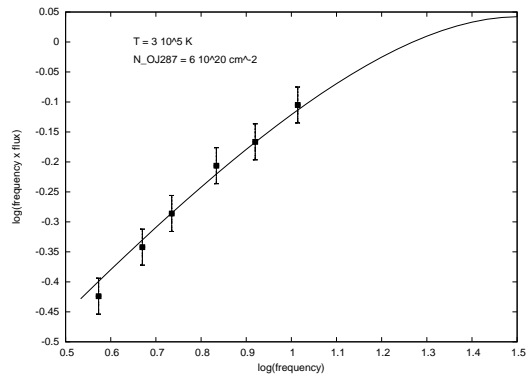} 
 \caption{The optical - UV spectrum of OJ287 during the 2005 outburst. Data points are based on Ref. 25. The solid line is the bremsstrahlung fit,  as predicted in Ref. 9. The observational points are corrected for the internal extinction in OJ287, taken from Ref. 35, and the assumed Galactic extinction law (Ref. 37). The standard extinction in our Galaxy is also taken into account.}
\end{figure}
We may also calculate the mass of the primary. Its value, $1.84\times10^{10}$ solar mass, seemed rather high when it was first calculated, but subsequent work on black hole mass functions now places it among the fairly common upper mass range black holes (common in the same sense as O-type stars are common among main sequence stars, see Refs. 26 - 29). This mass value places OJ287 right on the mean correlation of the black hole mass - host galaxy K-magnitude relation, with $M_K\sim-28.9$ (Refs. 30, 31).

In Ref. 12 it is claimed that OJ287 is significantly, slightly more than one standard deviation, off the mean correlation. Based on this, the authors state that the measurement in Ref. 30 ``is most likely spurious''. The reason for the one standard deviation offset in Ref. 12 may be traced to an incorrect way of transforming optical magnitudes to K-magnitudes. The process of transforming the R-band measurement of the host galaxy magnitude (Refs. 32 -34) to the K-band is composed of several steps, each containing its associated uncertainties. One has to have a theory of the stellar composition of the host galaxy (not trivial for a merger) and of its (passive) cosmological evolution. One has to measure the neutral hydrogen column density of the host galaxy, and then to transform it to the extinction in R-band. For the hydrogen column density there exists a measurement in Ref. 35, albeit with large error bars, while for the extinction curves large variations from galaxy to galaxy have been found (Ref. 36). As a result of these large uncertainties, one can safely say that the R-band measurements of the magnitude of the host galaxy are consistent with the direct measurement in the K-band (which does not require the above mentioned transformations), and that the black hole mass - host galaxy K-magnitude correlation holds in OJ287 within measurement errors. In any case, a displacement by one standard deviation from the mean correlation cannot be used as an argument for the correctness or otherwise of a single point in a correlation diagram.

\begin{figure}
\includegraphics[width=7cm,angle=0]{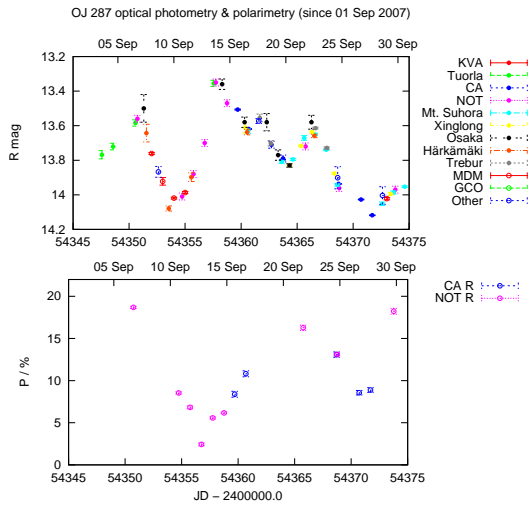} 
 \caption{The optical light curve of OJ287 during the 2007 September outburst. The upper panel shows the measured magnitudes, while the lower panel shows the percentage polarization. Data points are published in Ref. 22.}
\end{figure} 
\subsection{Nature of radiation at  the 2005 outburst}

Impact outbursts are expected to consist of bremsstrahlung radiation, and thus the optical polarization of OJ287 should go down during them. As mentioned above,  polarization information for the basic 2005 outburst is not available. However, bremsstrahlung may also be recognized by its spectrum, and this is the part of the campaign that was successfully carried out.

\begin{figure}
\includegraphics[width=5cm,angle=270]{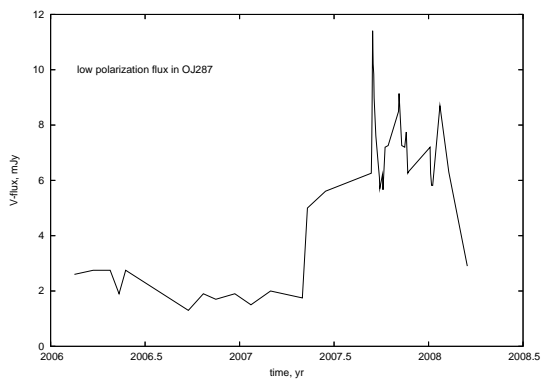} 
 \caption{The optical light curve of OJ287 during 2006-2008. Only low polarization (less than $10\%$) data are shown; they are based on ref. 24.}
\end{figure}

We had XMM-Newton observations both before the 2005 outburst (2005 April), and during the outburst (2005 November 3-4). Fortunately, the November observation happened at the time when the source was at its basic outburst level, in between two secondary bursts. Thus we would expect to see an additional pure bremsstrahlung spectrum above the underlying synchrotron power-law. A preliminary report of these observations has appeared in Ref. 25, and a more detailed report is under preparation.

\begin{figure}
\includegraphics[width=5cm,angle=270]{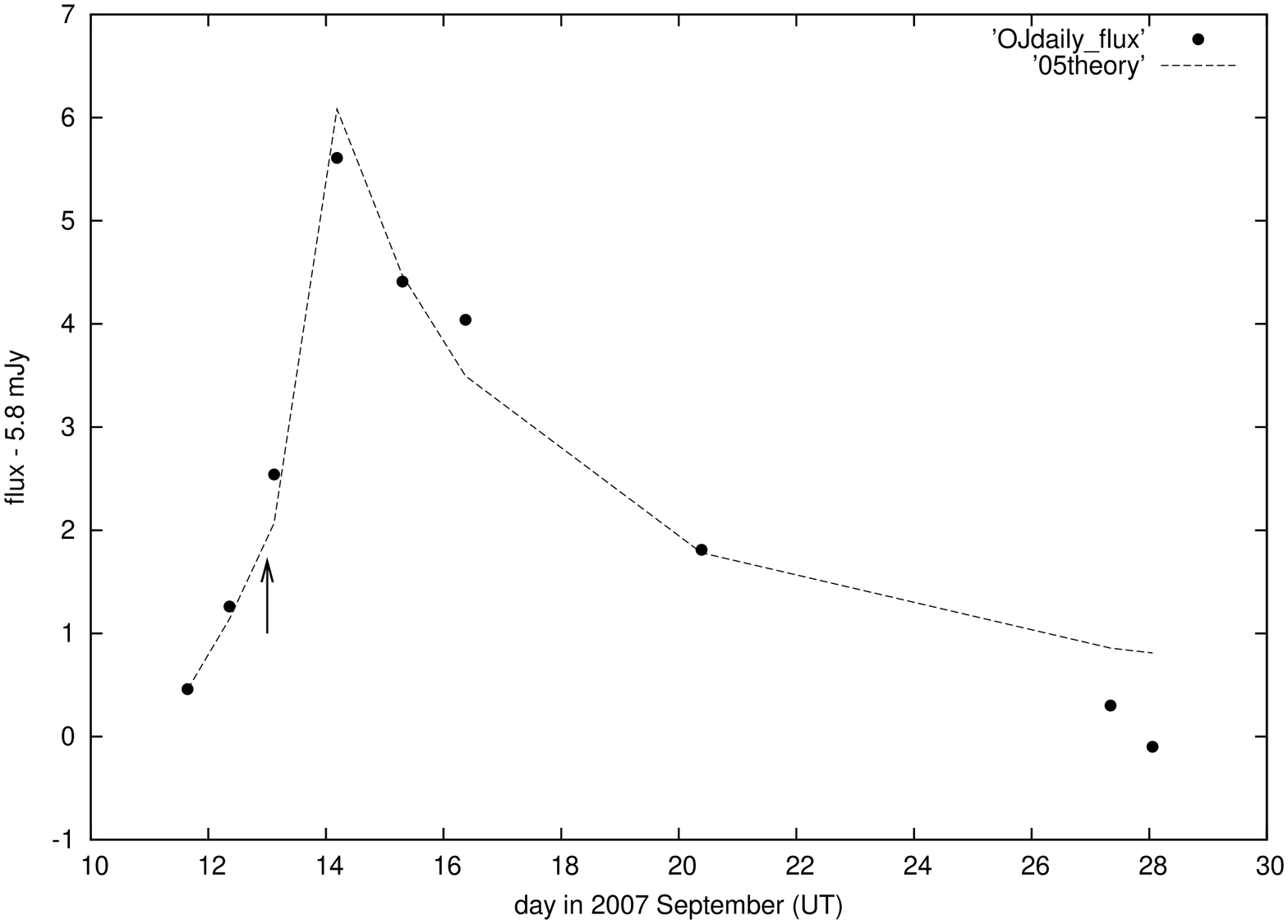} 
 \caption{The optical light curve of OJ287 during the 2007 September outburst. Only low polarization (less than $10\%$) data points are shown. The data points are based on Ref. 22. The dashed line is the theoretical fit based on Ref 9. The arrow points to September 13.0, the predicted time of origin of the rapid flux rise.}
\end{figure} 

In Figure 3 we show the difference between the 2005 November flux and the 2005 April flux. The values have been corrected for the Galactic extinction and for the extinction in the OJ287 host galaxy. For the latter, we use the measuments in Ref. 35 and the standard Galactic extinction curve (Ref. 37). The solid line shows the bremsstrahlung spectrum at the expected temperature of $3\times 10^{5~\circ}$K. Note that a raised synchrotron spectrum, as one might have expected in some other theories, would have a downward slope toward higher frequencies, and it is entirely inconsistent with observations. Incidentally, the effect of the extinction in the host galaxy of OJ287 is such that it causes an apparent spectral break in the normal (non-outburst) spectrum in the optical region, while it really happens at the AGN source somewhere in the UV (Ref. 35).

\subsection{Timing and nature of the 2007 September 13 outburst}

The 2007 September 13 outburst was an observational challenge, as the source was visible only for a short period of time in the morning sky just before the sunrise. Therefore a coordinated effort was made starting with observations in Japan, then moving to China, and finally to central and western Europe. A crucial role was played by the NOT telescope in the Canary Islands and by the Calar Alto telescope in mainland Spain, which were able to make polarization observations. The observed points with estimated error bars are given in Figure 4, where the contributions by participating observatories are shown by colour codes.

A comparison of the two panels shows immediately that there were two kinds of outbursts in 2007 September. Three outbursts were highly polarised, with the degree of polarization above $15\%$, while the biggest outburst had polarization below $10\%$. Thus it is not difficult to decide which was the expected bremsstrahlung event. Later in the year there were more highly polarized outbursts, but if we look at the light curve composed of low polarization states only (Figure 5), the September 13 outburst clearly stands out.

In Figure 6 we look at the low polarization light curve in more detail around the September 13 event. A theoretical light curve is also drawn, and an arrow points to the expected moment of the beginning of the sharp flux rise. We see that the observed flux rise coincides within 6 hours with the expected time. The accuracy is about the same as we were able to predict the return of Halley's comet with in 1986!

\subsection{2007 - 2010 outbursts}

Ref. 18 gave a detailed prediction of the whole light curve of OJ287 during the campaign period (Figure 7). In addition to the two impact outbursts, it was expected that the tidal forcing mechanism of Sillanp\"a\"a et al. [Ref. 5] would raise the general level of activity of OJ287, starting from the spring of 2007 and continuing until the spring of 2009. The detailed structure of minor bursts in Figure 7 is immaterial, since it is due to Poisson noise in a simulation with a finite number of disk particles. This prediction is best compared with the low-polarization light curve of Figure 5. In general outline OJ287 behaved just as expected, except that the optical flux declined fast in the spring of 2008, sooner than we would have thought.
 
\begin{figure}
\includegraphics[width=5cm,angle=270]{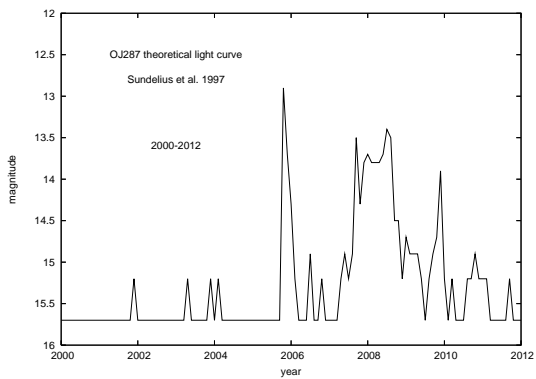} 
 \caption{The predicted optical light curve of OJ287 during 2000-2012. The data is based on Ref. 18.}
\end{figure} 

At this point we may remind the reader that Sillanp\"a\"a et al. [Ref. 5] also interpreted some sharp flux decreases as ``eclipses''. At these times the secondary may move across our line of sight, between us and the AGN optical continuum source. One such event was predicted in 2008 [Ref. 9], but it is not included in the light curve prediction of Figure 7. However, if we take the differential of observed minus predicted flux (Figure 8), the eclipse-like feature becomes quite obvious. The two previous ``eclipses'' in the same sequence occurred in 1989 [Ref. 38] and in 1998 [Ref. 39]. The astrophysical reason for the eclipses could be gravitational deflection of the jet stream by the secondary, extinction in gas clouds circling the secondary, or something else.

Figure 7 shows also a prominent outburst at the end of 2009. It also came as expected [Ref. 40].

In our model the accretion disk is optically thick but geometrically thin, it possesses a strong magnetic field [Ref. 41] and the disk is connected to the jet by magnetic field lines [Ref. 42]. Ref. 12 presents an entirely different model which they then strongly criticise, and finally try to make a case for quasi-periodic oscillations in an accretion disk of a single black hole. It is shown in Ref. 40 that the probability that such a model would explain the good match between the theory and observations is less than one in part in $10^8$, not to mention that the other ``smoking gun'' observations also remain unexplained in such a model. Actually, there is no evidence presented in favour of a single black hole model in Ref. 12, while the criticism of a binary model is misdirected and consists of a number of incorrect statements. 

\begin{figure}
\includegraphics[width=5cm,angle=270]{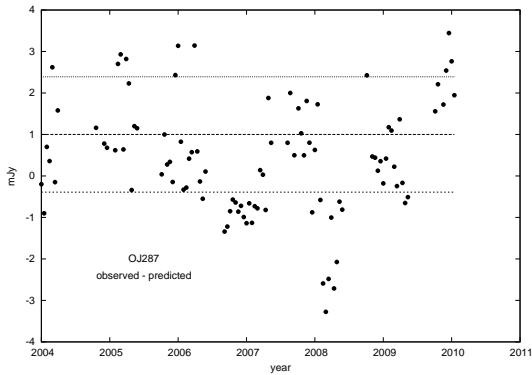} 
 \caption{The observed optical flux of OJ287 during 2004-2010 minus the prediction in Ref. 18. The scatter is 1.4 mJy, and the only significant deviation from the prediction occurs in the spring of 2008 which suggests an eclipse.}
\end{figure} 

\subsection{Turning jet}

The accretion disk as a whole is also affected by the companion in its 12 year orbit. On the other hand, in our model the jet and the disk are connected. Thus the jet direction should be strongly influenced by the companion. 

There are three periodicities that could be expected to show up: the 12 yr orbital cycle, the 120 yr precession cycle (or half of it due to symmetry) and the Kozai cycle [Ref. 43], which also happens to be 120 yr. The 12 yr orbital cycle produces the tidal enhancements in accretion flow, as postulated by Sillanp\"a\"a et al. [Ref. 5], but in addition this enhancement can be stronger or weaker depending on where we are in the precession cycle. These two tidal effects pretty much explain the overall appearance of the light curve [Ref. 18]. In addition, there is a modulation in the long term base emission level (unexplained by the tidal enhancement) which is in tune with the Kozai cycle. This cycle may also appear in polarization data [44]. The jet orientation is delayed relative to the disk wobble. Theoretically the delay should be of the order of ten years; fitting with the optical data gives the best fit with a 13 yr delay.

The jet wobble shows up in observations in several ways. First, the mean angle of optical polarization varies. The binary model predicts, among other things, a quick change in the optical polarization angle by nearly $90^{\circ}$ around 1995, which was observed [Ref. 12]. In radio, we should see a similar rapid change in the position angle of the parsec scale jet. Depending on the actual value of the delay in radio jet orientation, the change could already be under way (Figure 9), or it may be delayed by another 12 cycle (Figure 10, Ref. 45). There are recent observations which suggest the first alternative [Ref. 46], but the interpretation of these observations is not yet clear-cut.

There are also longer periods that are expected in the binary model: the period of the black hole spin (about 1300 yr, Refs. 47,48) which may show up in the structure of the megaparsec scale jet [Ref. 49]. Also the time scale of the binary settling in the nucleus of OJ287 after a merger of two galaxies, about $10^8$yr [Ref. 50], may be connected with the overall curvature of the magaparsec jet. In the shorter time scales, the half-period of the last stable orbit around the Kerr black hole of $\sim50$ days may also show up in observations [Ref.51].
 
\begin{figure}
\includegraphics[width=5cm,angle=270]{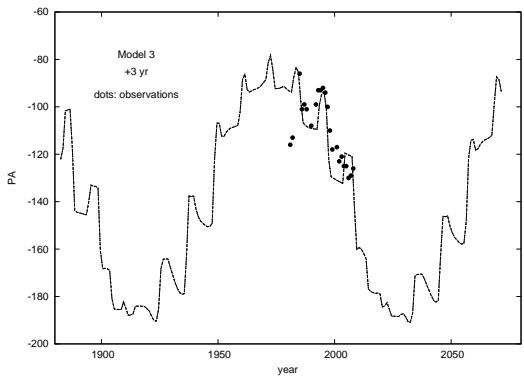} 
 \caption{The observed position angle of the radio jet of OJ287 (points) compared with the binary model, with a 3 year response time of the jet orientation changes.}
\end{figure} 
\begin{figure}
\includegraphics[width=5cm,angle=270]{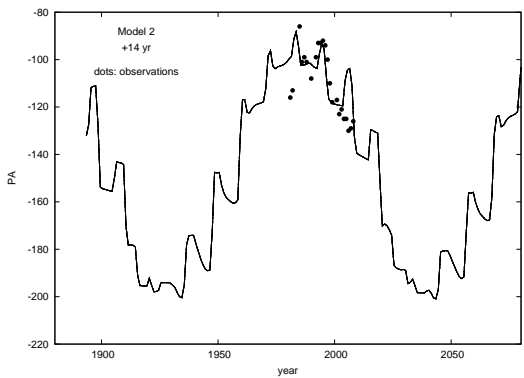} 
 \caption{The observed position angle of the radio jet of OJ287 (points) compared with the binary model, with a 14 year response time of the jet orientation changes.}
\end{figure}
 
\section{Testing General Relativity}

Using the OJ287 binary, we may test the idea that the central body is actually a black hole. One of the most important characteristics of a black hole is that it must satisfy the so called no-hair theorem or theorems (Refs. 52-57). A  practical test was suggested in Refs. 58, 59. In this test the quadrupole moment $Q$ of the spinning body is measured. If the spin of the body is $S$ and its mass is $M$, we determine the value of $q$ in
\begin{eqnarray}
Q = -q \frac{S^2}{Mc^2}.
\end{eqnarray}
For black holes $q=1$, for neutron stars and other possible bosonic structures $q > 2$ (Refs. 60, 61)

  We calculate the two-body orbit using the third Post-Newtonian (3PN) order orbital dynamics, which includes the leading order 
general relativistic, classical spin-orbit and radiation reaction effects (Refs. 62-64). 

The 3PN-accurate equations of motion can be written schematically as 
\begin{eqnarray} 
\ddot { {\vek x}} \equiv 
\frac{d^2  {\vek x}} { dt^2} &=& 
\ddot { {\vek x}}_{0} + \ddot { {\vek x}}_{1PN} \nonumber 
+ \ddot { {\vek x}}_{SO}+\ddot { {\vek x }}_{Q}\\ 
&& + \ddot { {\vek x}}_{2PN} +  \ddot { {\vek x}}_{2.5PN}
+ \ddot { {\vek x}}_{3PN} \,,  
\end{eqnarray} 
where ${\vek x} = {\vek x}_1 - {\vek x}_2 $ stands for the 
center-of-mass 
relative separation vector between the black holes with masses $m_1$ and $m_2$ and 
$  \ddot { {\vek x}}_{0}  $ represents the Newtonian acceleration given by  
$ \ddot { {\vek x}}_{0} = -\frac{ G\, m}{ r^3 } \, {\vek x} $; $m= m_1 + m_2$ and $ r = | {\vek x} |$. 
The PN contributions occurring at the conservative 1PN, 2PN, 3PN and the reactive 2.5PN orders, denoted 
by $\ddot { {\vek x}}_{1PN}$, $\ddot { {\vek x}}_{2PN}$, $\ddot { {\vek x}}_{3PN}$ and 
$\ddot { {\vek x}}_{2.5PN}$ respectively, are non-spin by nature, while $ \ddot { {\vek x}}_{SO}$ is the spin-orbit term of the order 1.5PN.

The quadrupole-monopole interaction term $\ddot { {\vek x}}_Q $, entering at the 2PN order, reads
\begin{eqnarray} 
\ddot { {\vek x}}_Q & =- q \, \chi^2\, 
\frac{3\, G^3\, m_1^2 m}{2\, c^4\, r^4}
\, \biggl \{ 
\biggl [ 5(\vek n\cdot \vek s_1)^2 
-1 \biggr ] {\vek n}
-2(\vek n\cdot \vek s_1) {\vek s_1} \biggr \}, 
\end{eqnarray} 
where parameter $q$, whose value is $1$ in general relativity, is introduced to test the black hole `no-hair'   
theorem. The Kerr parameter $\chi$ and the unit vector 
${\vek s}_1$ define the spin of the primary black hole by the relation 
${\vek S}_1 = G\, m_1^2 \, \chi\, {\vek s}_1/c$ 
and $\chi$ is allowed to take values between $0$ and $1$ in general relativity. The unit vector {\vek n} is along the direction of {\vek x}.
 
In Figure 11 we show the distribution of $q$-values allowed by ``good'' orbits. By ``good'' we mean an orbit which gives the correct timing of 9 outbursts within the range of measurement accuracy. Obviously the range of timing at each of the 9 outbursts means that a set of solution orbits are possible. Here we have used a representative set of such orbits.

We note that the distribution peaks at $q=1$, thus confirming the no-hair theorem. It is also the first test of general relativity that has been performed at higher than the 1.5 Post-Newtonian order. Thus it forms a milestone in our study of the correct theory of gravitation.
 
\begin{figure}
\includegraphics[width=5cm,angle=270]{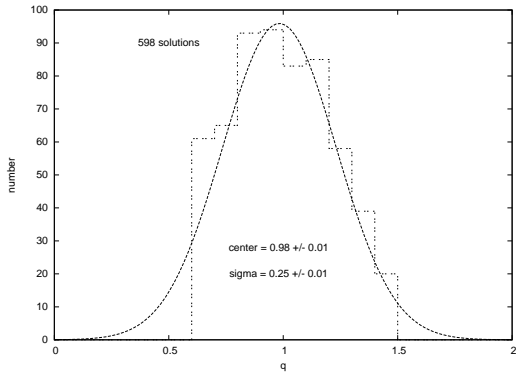} 
 \caption{The distribution of the test parameter $q$ among 598 solutions of the orbit. The result is consistent with General Relativity ($q=1$), and excludes the cases of no relativistic spin-orbit coupling ($q=0$) and of a material body ($i.e.$ not a black hole, $q$ greater than 2).}
\end{figure} 
\section{Conclusions}

Prior to the 2005 - 2010 multiwavelength campaign there were several ideas about the nature of OJ287. Fortunately, these models made completely different predictions about the behaviour of OJ287 during these years. One of the key differences was the timing of the first outburst: the precessing binary model, initially proposed by Sillanp\"a\"a et al. [Ref. 5] with subsequent refinements [Refs. 9, 10, 18] predicted the outburst in October 2005, the other models in October 2006. The result of a scientific enquiry is seldom as clear-cut as this: the outburst came within one week of the time expected in the precessing binary model and its spectrum agreed with the bremsstrahlung spectrum at the predetermined temperature. The prediction for the second outburst turned out to be accurate within 6 hours, and the lack of polarization again suggested strongly the bremsstrahlung origin. All flux values predicted for this period turned out to be accurate with the standard scatter of 1.4 mJy, which is only ten percent of the variability range in optical. The only exception occurred in 2008; however, this was the time when lower flux values were expected due to an ``eclipse''. We have ``eclipse'' in quotation marks, as the reason for the sudden fade at the time when the secondary passes through our line of sight is not known.

The optical variability data specifies the binary model except for the exact direction of the jet relative to our line of sight. However, the resolved parsec scale radio jet allows us to get a handle on this parameter, too. The remaining unknown is the delay between the wobble of the accretion disk, due to the effect of the secondary, and the reorientation of the jet in the sky. A major reorientation may already have started, or it may come after one orbital period, depending on the details of the jet/disk connection [Ref. 65].

The success of the binary model has encouraged us in using it to test theories of gravitation. Any theory which can be presented as Newton's law plus Post-Newtonian terms may be studied, as different laws of gravitation produce different impact times on the accretion disk. We have used the Post-Newtonian terms of general relativity up to third order, and have found that the orbit solutions agree with the theory. Our test parameter $q$ should have the value of exactly 1, and indeed the possible solutions cluster around this value. The parameter values $q=0$ or $q=2$ can be rejected at the 4 standard deviation level at present. This is the first time that it has been possible to study general relativity at higher than the 1.5PN order [Ref. 66].

\section{Acknowledgement}

We would like to thank all the participants in this campaign for the extraordinary amout of work dedicated to solving the riddle of OJ287. In particular, we thank Kari Nilsson and Stefano Ciprini, who have put together and harmonized huge amounts of data. Rene Hudec has made an invaluable contribution in collecting historical data, Tapio Pursimo and Leo Takalo were responsible for the key polarization observations at NOT, and Seppo Mikkola, Harry Lehto and Achamveedu Gopakumar have been key persons in turning the data into a viable binary black hole model.


\begin{thebibliography}{20}

\bibitem{}Haarala, S., Korhonen, T., Sillanp\"a\"a, A. \& Salonen, E., IAU Circ., 3764, 3 (1983) 

\bibitem{}Sillanp\"a\"a, A., et al., Astron. \& Astrophys., 147, 67 (1985) 

\bibitem{}Smith, P. S., Balonek, T. J., Elston, R., \& Heckert, P. A., Astrophys. J. Suppl., 64, 459 (1987) 

\bibitem{}Valtaoja, L., Sillanp\"a\"a, A. \& Valtaoja, E., Astron. \& Astrophys., 184, 57 (1987)

\bibitem{}Sillanp\"a\"a, A., Haarala, S., Valtonen, M. J., Sundelius, B. \& Byrd, G. G., Astrophys. J., 325, 628 (1988)

\bibitem{}Fiorucci, M., et al., IAU Circ., 6104, 2 (1994) 

\bibitem{}Sillanp\"a\"a, A. et al., Astron. \& Astrophys., 305, L17 (1996) 

\bibitem{}Valtonen, M. J., Workshop on Two Years of Intensive Monitoring of OJ 287 and 3C66A, Proceedings of the meeting held in Oxford, England, 11-14 September, 1995. Tuorla Observatory Reports, Informo No. 176. Edited by Leo O. Takalo, Tuorla Observatory, University of Turku, p.64 (1996) 

\bibitem{}Lehto, H. J. \& Valtonen, M. J., Astrophys. J., 460, 207 (1996) 

\bibitem{}Sundelius, B., Wahde, M., Lehto, H. J. \& Valtonen, M. J., Blazar continuum variability Astronomical Society of the Pacific Conference Series 110, Proceedings of an international workshop held at Florida International University, Miami, Florida, USA, 4-7 February 1996, San Francisco: Astronomical Society Pacific, edited by H. Richard Miller, James R. Webb, and John C. Noble, p.99 (1996) 

\bibitem{}Sillanp\"a\"a, A. et al., Astron. \& Astrophys., 315, L13 (1996)

\bibitem{} Villforth, C. et al., Mon. Not. RAS, 402, 2087 (2010)  

\bibitem{}Igumenshchev, I. V. \& Abramowicz, M. A., Mon. Not. RAS, 303, 309 (1999) 

\bibitem{}Katz, J. I., Astrophys. J., 478, 527 (1997)

\bibitem{}Villata, M., Raiteri, C. M., Sillanp\"a\"a, A. \& Takalo, L. O., Mon. Not. RAS, 293, L13 (1998)

\bibitem{}Valtaoja, E., Ter\"asranta, H., Tornikoski, M., Sillanp\"a\"a, A., Aller, M. F., Aller, H. D. \& Hughes, P. A., Astrophys. J., 531, 744 (2000)

\bibitem{}Kidger, M. R., Astron. J., 119, 2053 (2000)

\bibitem{}Sundelius, B., Wahde, M., Lehto, H. J. \& Valtonen, M. J., Astrophys. J., 484, 180 (1997)

\bibitem{}Valtonen, M. J., Astrophys. J., 659, 1074 (2007)

\bibitem{}Valtonen, M. J. The Nuclear Region, Host Galaxy and Environment of Active Galaxies: Proceedings of a Meeting to celebrate the 60th Birthday of Deborah Dultzin-Hacyan, Huatulco, Mexico, 18-20 Apr 2007 (Eds. Erika Benítez, Irene Cruz-González, \& Yair Krongold) Revista Mexicana de Astronomía y Astrofísica (Serie de Conferencias) Vol. 32, 22 (2008)

\bibitem{}Valtonen, M. J. et al., Astrophys. J., 646, 36 (2006)

\bibitem{}Valtonen, M. J. et al., Nature, 452, 851 (2008)

\bibitem{}Valtonen, M., Kidger, M., Lehto, H. \& Poyner, G., Astron. \& Astrophys., 477, 407 (2008)

\bibitem{}Valtonen, M. J. et al., Astrophys. J., 698, 781-785 (2009)
	
\bibitem{}Ciprini, S. et al., Mem. Soc. Astronomica Italiana, 78, 741 (2007)
	
\bibitem{}Ghisellini, G. et al., Mon. Not. RAS, 399, L24 (2009)

\bibitem{}Sijacki, D., Springel, V. \& Haehnelt, M. G., Mon. Not. RAS, 400, 100 (2009)

\bibitem{}Kelly, B. C. et al., Astrophys. J., 719, 1315 (2010)

\bibitem{}Trakhtenbrot, B., Netzer, H., Lira, P. \& Shemmer, O., Astrophys. J., 730, 7 (2011)

\bibitem{}Wright, S. C., McHardy, I. M. \& Abraham, R. G., Mon. Not. RAS, 295, 799 (1998)

\bibitem{}Kormendy, J. \& Bender, R., Nature, 469, 377 (2011).

\bibitem{}Hutchings, J. B., Astrophys. J., 320, 122 (1987)

\bibitem{}Heidt, J. et al., Astron. \& Astrophys., 352, L11 (1999) 

\bibitem{}Pursimo, T. et al., Astron. \& Astrophys., 381, 810 (2002) 

\bibitem{}Ghosh, K. K. \& Soundararajaperumal, S., Astrophys. J. Suppl., 100, 37 (1995)

\bibitem{}Falco, E. E. et al., Astrophys. J., 523, 617 (1999)

\bibitem{}Draine, B. T., Ann. Rev. Astron. \& Astrophys., 41, 241 (2003)

\bibitem{}Takalo, L. O. et al., Astron. \& Astrophys. Suppl., 83, 459 (1990).

\bibitem{}Valtonen, M. J., Lehto, H. J. \& Pietil\"a, H., Astron. \& Astrophys., 342, L29 (1999)
	
\bibitem{}Valtonen, M. J., Lehto, H. J., Takalo, L. O. \& Sillanp\"a\"a, A., Astrophys. J., 729, 33 (2011)
 
\bibitem{}Sakimoto, P. J. \& Coroniti, F. V., Astrophys. J., 247, 19 (1981)

\bibitem{}Turner, N. J., Bodenheimer, P. \& Rozyczka, M., Astrophys. J., 524, 129 (1999)

\bibitem{}Innanen, K. A., Zheng, J. Q., Mikkola, S. \& Valtonen, M. J., Astron. J., 113, 1915 (1997)

\bibitem{}Sillanp\"a\"a, A., Astron. \& Astrophys., 247, 11 (1991)
	
\bibitem{}Valtonen, M. J., Savolainen, T. \& Wiik, K., Jets at all Scales, Proceedings of the International Astronomical Union, IAU Symposium, Volume 275, p. 275 (2011)

\bibitem{}Agudo, I. et al., Astrophys. J., 726, L13 (2011)

\bibitem{}Valtonen, M. J. et al., Astrophys. J., 709, 725 (2010)

\bibitem{}Valtonen, M. J. et al., Celestial Mechanics and Dynamical Astronomy, 106, 235 (2010)
	
\bibitem{}Marscher, A. P. \& Jorstad, S. G., Astrophys. J., 729, 26 (2011)

\bibitem{}Iwasawa, M., An, S., Matsubayashi, T., Funato, Y. \& Makino, J., Astrophys. J., 731, L9 (2011)

\bibitem{}Wu, J. et al., Astron. J., 132, 1256 (2006)

\bibitem{} Israel, W., Phys.Rev. 164, 1776 (1967)

\bibitem{} Israel, W., Commun. Math.Phys. 8, 245 (1968)

\bibitem{}  Carter, B., Phys.Rev.Lett. 26, 331 (1970)

\bibitem{} Hawking, S.W., Phys.Rev.Lett. 26, 1344 (1971)

\bibitem{} Hawking, S.W., Commun.Math.Phys. 25, 152 (1972)

\bibitem{} Misner, C.W., Thorne, K.S. \& Wheeler, J.A., Gravitation, W.H.Freeman \& Co, New York, p. 876 (1973)

\bibitem{} Thorne, K.S., Rev.Mod.Phys. D, 31, 1815 (1980)

\bibitem{} Thorne, K.S., Price, R.M. \& Macdonald, D.A., in Black Holes: The Membrane Paradigm, Yale Univ. Press, New Haven (1986)

\bibitem{} Wex, N. \& Kopeikin, S.M. 1999, Astrophys. J., 514, 388 (1999)

\bibitem{} Will, C. M., Astrophys. J., 674, L25 (2008)

\bibitem{}Barker, B. M. \& O'Connell, R. F., PhRvD, 12, 329 (1975)

\bibitem{} Damour, T. 1982, C.R. Acad. Sci. Paris 294, (II), 1355 (1982)

\bibitem{} Kidder, L. E., PhRvD, 52, 821 (1995)

\bibitem{} Valtonen, M.J., Villforth, C. \& Wiik, K., Mon. Not. RAS, in press, arXiv1111.1539 (2011)

\bibitem{} Valtonen, M. J., Mikkola, S., Lehto, H.J., Gopakumar, A., Hudec, R. \& Polednikova, J., Astrophys. J., 742, 22 (2011)

\end{thebibliography}
\end{document}